# ORCA: Enabling an Owner-centric and Data-driven Management Paradigm for Future Heterogeneous Edge-IoT Systems

Jianli Pan, Jianyu Wang, Ismail AlQerm, Yuanni Liu, and Zhicheng Yang

*Abstract*—**Integrating Internet of Things (IoT) and edge computing for "Edge-IoT" systems, converged with machine intelligence, has the potentials of enabling a wide range of applications in smart homes, factories and cities. Edge-IoT can connect many diverse devices and the IoT asset owners can run heterogeneous IoT systems supported by various vendors or service providers (SPs), using either cloud or local edge computing (or both) for resource assistance. The existing methods typically manage the systems as separate vertical "silos", or in a vendor/SP-centric way, which suffers from significant challenges. In this paper, we present a novel owner-centric management paradigm named "ORCA" to address the gaps left by the owner-centric paradigm and empower the IoT assets owners to effectively identify and mitigate potential issues in their own network premises, regardless the vendors/SPs' situations. ORCA aims to be scalable and extensible in assisting IoT owners to perform intelligent management through a behavior-oriented and data-driven approach. ORCA is an ongoing project and the preliminary results indicate that it can significantly empower the IoT systems owners to better manage their IoT assets.**

## I. INTRODUCTION

The National Academy of Engineering (NAE) identified 14 grand challenges [1] our society faces, including virtual reality, health informatics, secure cyberspace, clean water, and urban infrastructure. They can be directly benefited by integrating Artificial Intelligence (AI), machine intelligence, Internet of Things (IoT), edge computing, and 5G to closely work for the citizens, businesses, and the whole society. Future smart homes, factories, communities and cities will also be empowered. We envision a future "Edge-IoT" environment [2] converged with machine intelligence and data-driven approaches to better serve the people and businesses. Edge-IoT can connect massive numbers of smart devices, and IoT asset owners can run heterogeneous IoT systems supported by various vendors or Service Providers' (SPs) platforms, and can use either cloud or local edge computing (or both) for resource assistance. However, it is a significant challenge to scalably and effectively manage such a dramatic number and variety of devices, and heterogeneous Edge-IoT systems. Poor management partially contributes to the large-scale botnet attacks and significant financial loss [3]. Specifically, current

J. Pan, J. Wang, and I. AlQerm are with the Department of Computer Science, University of Missouri, St. Louis, MO, USA 63121. E-mail: pan, jwgxc, alqermi@umsl.edu. Y. Liu is with ChongQing University of Posts and Telecommunications, ChongQing, China. Z. Yang is with PAII Inc, USA.

IoT systems are typically managed by different vendors/SPs as separate vertical "silos" [4]. The vendor/SP-centric management overly relies on vendors' uneven capabilities, and lacks transparency and cross-subsystem insights for the owners. The owners are also at risk of losing basic management capabilities when vendors/SPs run into abnormal situations or go out of business.

In this paper, we envision building a novel owner-centric management paradigm to fill the existing gaps and empower the IoT owners to manage across subsystems that the vendors/SPs are currently not able to do. The owners are in the most capable and suitable position of in-premises edge networks (not cloud) to effectively identify and mitigate potential issues. The significance of the new paradigm is multifold. First, it empowers owners to manage diverse devices and complex behavior, and can greatly reduce the financial loss due to management failure. Second, it enables the owners to manage across subsystems when separate silos are not fully interoperable and standardization falls behind. Third, it enables owners to continue managing their assets even when vendors/SPs stop support or are out of business.

However, significant technical barriers exist to enable this new paradigm. First, existing methods in industry and academia either only manage small device variety in dedicated silos and only consider simple behavior, or use limited data source such as network traffic. In such tasks, simple statistic or machine learning methods suffice and they can afford relatively expensive samples labeling. But the owners may have to manage a large device variety and complex behavior patterns, and large-scale sample labeling also becomes economically infeasible. Second, the existing methods do not account for scalability and extensibility to accommodate owners' growing management interests. Target behavior may also have different complexity and the modeling approaches should be customizable to balance between performance and cost. Third, the existing data-driven management methods in both industry and academia focus on small scopes, and it lacks a holistic full-cycle data-driven approach to empower the IoT owners across the whole management cycle of "observing, synthesizing, and responding".

In this paper, we aim to present a scalable and extensible owner-centric management framework named ORCA to assist IoT owners to perform intelligent management for diverse devices and heterogeneous Edge-IoT systems through a full-cycle data-driven approach. Specifically, ORCA holistically addresses the above technical barriers via a series of unique designs and contributions. First, it adopts a unique behavior-oriented and data-driven approach to allow owners to model complex behavior of diverse devices and heterogeneous Edge-IoT systems utilizing various data sources. Second, ORCA allows the owners to scalably and extensibly define and deploy multi-level observable "behavior" models (output as "insights"), identify suitable modeling approaches based on behavior complexity and data features, and balance performance and cost. Third, ORCA provides full-cycle customized data-driven toolsets for the IoT owners to model device behavior, synthesize cross-silo group behavior, and make intelligent management decisions without being required to have deep technical expertise. Fourth, ORCA runs at edge premises instead of in cloud, avoids excessive data transmission and delay, and can manage when offline. It is run by owners, independent of the existing functions in silos, and can continue managing when vendors stop supports or even are out of business.

TABLE I: Example Edge-IoT systems and characteristics.

| Devices and Applications | Data type | Priority (1-4: H to L) | Computing Intensiveness | Data Intensiveness | Latency Sensitivity |
|---|---|---|---|---|---|
| Emergency real-time response (e.g. gunshot detection ) | video/audio | 1 | high | high | high |
| VR/AR related applications | video | 2 or 3 | high | high | high |
| Home voice assistant | audio | 2 | medium | medium/low | high |
| Cognitive assistance | video/audio | 2 or 3 | high/medium | high/medium | medium |
| Building access face detection | video | 3 | medium | medium/low | medium |
| Personal Identification | audio/image/text | 3 | medium/low | medium/low | medium |
| Home health monitoring | text | 2 | low | low | low |
| Common smart home devices | text/audio | 4 | medium/low | low | low |
| Low-level sensors | text | 4 | low | low | low |

The rest of the paper is organized as follows. Section II is related work and the current vendor/SP-centric paradigm. The ORCA rationale is in Section III. Section IV is the ORCA architecture and the data-driven 3-step IoT management. Some preliminary evaluation and discussions are in Section V. The conclusions follow in Section VI.

## II. CURRENT IoT MANAGEMENT AND VENDOR/SP-CENTRIC PARADIGM

In this section, we discuss the current related work and issues on IoT management.

### A. Related Work

IoT management has been studied on individual aspects such as trust, resource, energy/power, data, and privacy management. Management has also been closely tied to specialized devices in industrial factory machinery, power grid, water network, and supply chain. For example, Prognostics and Health Management (PHM) [5] for industrial machinery health diagnostics has been heavily researched, using the recent statistical and machine learning methods. Such industrial applications are recently moved to cloud and managed by vendors/SPs as separate "silos" [4]. Some examples include *Amazon AWS IoT*, *IBM Watson IoT*, *Google Cloud IoT*, and small vendors renting cloud to provide supports. Inside each "silo", machine learning based IoT analytics are performed for predictive maintenance, big-data inference, and anomaly identification. Mobile device management (MDM) [6] deals with smartphone management with limited types of devices and Operating Systems (OSs). Its goals and scopes are different from the IoT management. Another category of works use machine learning techniques over data traffic for device fingerprinting, behavior analysis, and intrusion detection. Typical commercial products include *Extreme IoT Defender*, *Zingbox*, and *Cisco Appdynamics*. However, using only traffic analysis is limited. The used machine learning methods require expensive labeling and are inadequate to model very diverse devices and growing management interests. In addition, "horizontal" efforts [7] aim for better interoperability among silos. Example efforts include standardization, industry alliances, IoT ontologies [8], and market convergence. Horizontal process is relatively slow, and by itself it cannot lead to owner-centric management.

### B. Various Management Modes and the Vendor/SP-centric Paradigm

Depending on the actual cases and business models, different management modes exist. The IoT systems can be managed either by owners themselves locally or by specialized vendors/SPs at edge or clouds. We consider two key factors: owners' characteristics and capabilities, and the vendors/SPs' expertise and capability. On one hand, various owners may have very different expertise and capability. The first category is that the owners run very dedicated applications such as Industry 4.0 factories, smart vehicle charging network, and camera-based security events detection. In these cases, the owners are either very capable and can manage all the specialized devices by themselves locally or at edge, or rely on very powerful vendors/SPs such as Google and Amazon to manage at edge or cloud. For these ideal cases, existing management methods may suffice. The second category are less ideal cases in that the common IoT owners are much less capable, and they barely have adequate expertise or tools to manage things all by themselves. On the other hand, various vendors/SPs that the owners rely on may also have very different capability. For example, they may range from powerful companies such as Amazon and Google, to the vendors of many cheap devices that barely provide any management.

Meanwhile, it has become common that IoT owners may own and run multiple systems in their networks, and they rely on various vendors/SPs to manage these systems separately. For example, a smart home owner may run NEST on Google cloud for thermostat, Amazon's Echo on AWS cloud for voice assistant, and some cheap OEM IP cameras from small vendors on rented cloud spaces. These subsystems are either managed by owners themselves or in a vendor/SP-centric way, i.e., by different device vendors or SPs vertically as separate "silos" [4] that typically do not share interfaces, data and insights. Within a foreseeable future, the IoT market will remain scattered with various sizes of vendors/SPs using different software and platforms. The standardization process remains relatively slow and the interoperability between different platforms is limited.

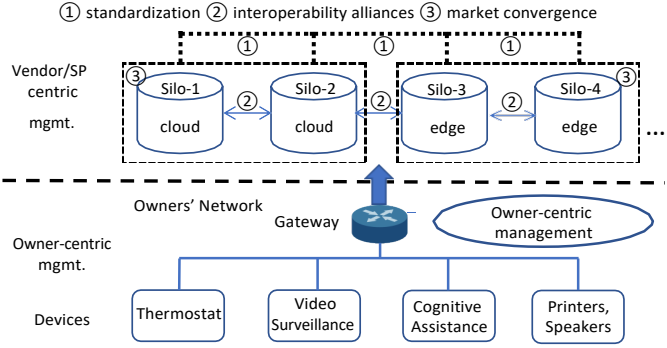

Fig. 1: Owner-centric vs. vendor/SP-centric paradigms.

In addition, a series of emerging trends will bring even more difficulties to the management. First, there are increasing numbers of devices and growing management interests. Second, the devices can be very diverse [9] in constrainedness, network types (wireline/wireless), protocols (Wi-Fi/Ethernet/5G), network patterns (P2P/P2MP/multi-hop), media types (text/audio/video) and characteristics (running modes, bandwidth, and response frequency). They may show a wide range of behavior patterns. Third, heterogeneous IoT systems are with various quality requirements. Some typical examples are shown in Table I, which includes those resource-intense and latency-sensitive applications (in light-gray shades). The existing management practices of the IoT systems have fallen short, which has been partially reflected in the wide-spreading large-scale botnet attacks and significant financial losses caused by millions of poorly managed smart IoT devices in recent years [3].

## III. Owner-centric Paradigm and Rationale

In this section, we will discuss the new paradigm and the designing rationale.

### A. Owner-centric vs. Vendor/SP-centric Paradigms

For the common owners, overly relying on the vendor/SP-centric paradigm may incur significant limitations. Thus, we propose a new owner-centric paradigm named ORCA to fill the existing gaps left by the vendor/SP-centric paradigm (not to replace the application logic of individual silos), and provide much-needed advanced designs to empower those vulnerable and incapable IoT owners to better manage various heterogeneous IoT systems. Specifically, ORCA addresses the challenges as follows. First, various vendors/SPs may have very uneven technical capabilities, and the powerful vendors/SPs cannot manage devices across silos. This may result in poor management of some devices (like some OEM IP cameras) by less capable vendors/SPs. They can become weak links and be compromised and used as springboard by the hackers to launch internal attacks. ORCA provides owners with basic management for these vulnerable devices, so that they will not be easily exploited by hackers. Second, the management functions of various vendors/SPs are typically separate and not transparent, and the owners generally have no way to gain cross-subsystem insights. For example, the owners may want to know whether the shared resource pool are being responsibly used by different subsystems, and whether devices abnormally interact with others managed by different vendors/SPs. ORCA will allow the owners to manage beyond silos and extract useful cross-system or group insights by performing better data analysis defined by the owners and serving their own objectives. The owners are in better position to judge how these groups of devices from different vendors can be managed to serve the owner purposes and fit with its current facilities. Third, when specific vendors/SPs experience temporary/permanent situations, stop support or go out of business, the owners are at the risk of losing basic management capabilities over their own assets. ORCA will continue providing basic management support even if the above situations occur and it will reduce management lapses that hackers can exploit. Fourth, instead of managing from cloud, ORCA manages at the owners' network premise, which does not incur large volume of data transmission and long delay, and does not require the owner network to be always online.

We compare the two paradigms in Fig. 1 via a simple smart home example with four silos including subsystems using cloud or edge. It also shows the three types of horizontal efforts including standardization, interoperability alliances and market convergence, and illustrates their relationship. Horizontal progress can potentially alleviate some of the challenges and help management with better data quality and availability, hence benefit ORCA's performance. However, horizontal progress does not automatically lead to owner-centric management. Even in a "perfect" world with full standardization and interoperability, and with fewer silos, the management is still in a vendor/SP-centric way. The horizontal vision will be a relatively long way to go.

### B. Behavior-oriented and Data-driven Approaches

We define "behavior" as the patterns that the IoT owners want to observe based on their extensible management interests of different levels [10]. 1) B1: device level interests such as hardware failure, software malfunction, and remaining lifetime. 2) B2: network level interests such as traffic patterns, unsafe connections, and botnet activities. 3) B3: cloud/edge interests such as requested resource, offloaded tasks, and response time. 4) B4: group and subsystem level interests such as how groups behave in B1 to B3. In ORCA, to generate device behavior "insights" in a timely manner and at suitable cost, specialized machine learning models are needed that are suitable to the behavior complexity and can appropriately balance between performance and cost for both model training and operation. The overall idea is illustrated in Fig. 2. The owner will be able to deploy behavior models flexibly and extensibly based on device types and their management interests (B1-B4). Moreover, ORCA is data and insights driven. The device-level behavior modeling results are used to synthesize group and system level behavior, and they are further used to assist owners to make intelligent management responses.

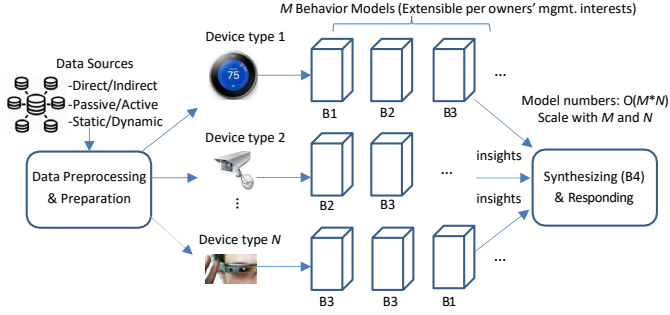

Fig. 2: Scalable and extensible behavior models.

## IV. ORCA ARCHITECTURE AND DATA-DRIVEN 3-STEP MANAGEMENT

In this section, we focus on the ORCA architecture and the proposed data-driven 3-step IoT management.

### A. ORCA Architecture Overview

The ORCA architecture is presented in Fig. 3. It works at the network edge and integrates the IoT device side and the resource side. The device side includes multiple types of devices and subsystems. The resource side consists of multiple edge servers comprising virtual machines, communication and computation resources. The acquired data from both sides go through quality improvement and the resulted data samples are used to train or retrain the models. The device behavior manager will profile the device behavior based on managers' interests, and decide the appropriate candidate models based on the data time dependency and the behavior complexity. The group behavior manager synthesizes device-level behavior into group-level insights by clustering, and group resource usage trends by Long Short Term Memory (LSTM) [13] based prediction. The intelligent response manager will utilize the insights from the device manager and group manager to make intelligent decisions in device-level predictive maintenance, and Quality of Experience (QoE) based intelligent edge resource allocation. Specifically, the device predictive maintenance module will further inspect the group outliers identified by the clustering and make behavior predictions using an online and lightweight OL-ARIMA [12] approach. It will then generate a device list for further maintenance. The resource allocation module aims to build a QoE model, takes various behavior insights as parameters, and allocates edge resources using a two-stage deep online learning method, with the goals of maximizing users' satisfaction and edge resource utilization, and encouraging good behavior.

### B. Data-driven 3-step Management

1) "Observing": Owner-centric Device-level Behavior Modeling: This step focuses on IoT owners' device manager role and aims to profile and model various behavior of diverse devices for heterogeneous IoT systems from an owner-centric perspective with growing management interests.

i) Behavior Profiling. We first profile multi-level new behavior targets, exploits various data sources, and maps the behavior targets to suitable models. The device behavior profiling workflow is shown in Fig. 4. The workflow inputs are the target behavior that the owners choose to model for a specific type of device, and the outputs are the trained models that have learned the patterns of device normal behavior. To jointly consider performance, cost and scalability, we formulate behavior modeling as one-class classification problem [11] in order to avoid the difficult and expensive labeling for anomalies of diverse devices and relatively low frequency for specific faults, and leverage the relatively ample "normal" behavior data to train the one-class classifiers. The behavior models will output the devices' current abnormal degree comparing with the normal or healthy condition. For example, if an IoT owner wants to define a behavior model for an IP camera in B2 category focusing on its traffic pattern, the model output will be a score measuring its abnormal degree. If the camera is having a software malfunction or a botnet attack, the model will output a score in an alarming range. Depending on behavior types (B1 to B3), various data sources related to multi-level features across IoT reference model are used to prepare training samples and dataset.

ii) Behavior Modeling. We then map the behavior target to the suitable behavior modeling methods based on two data attributes: time dependency and feature dimensionality, in order to balance between performance (speed/accuracy) and cost (training/operation). First, key behavior data may be either time series or non-time series data. Time series data are generated periodically at fixed intervals such as sensors data or network packets with natural time stamps, and the patterns may exist in the time dependency. Non-time series data have little or none time dependency between instances. Second, complex behavior's high feature dimensionality may cause high data sparsity, and require advanced modeling approaches and more training samples. Thus, to effectively learn high dimensional feature distributions in the one-class classifiers, we integrate the Generative Adversarial Network (GAN) [14] with Encoder-Decoder networks (GAN-ED) for non-time series behavior modeling, and integrate LSTM [13] with ED networks (LSTM-ED) for time-series behavior modeling. In the GAN-ED model, we incorporate an encoder network into the original GAN framework. The encoder advances its learning ability that compresses the inputs into low-dimensional feature vectors through the mutual training process with the generative and discriminative networks in GAN. In the LSTM-ED model, we employ LSTM neurons in the hidden layers for both the encoder and the decoder networks to learn the time dependencies between the input features. For simpler behavior with low dimensionality of the non-time series data, we develop a simpler method based on One-class Support Vector Machine (OC-SVM) for fast and less-costly modeling, and an agile and lightweight prediction method based on Multivariate Autoregressive Integrated Moving Average (MARIMA) for time series data.

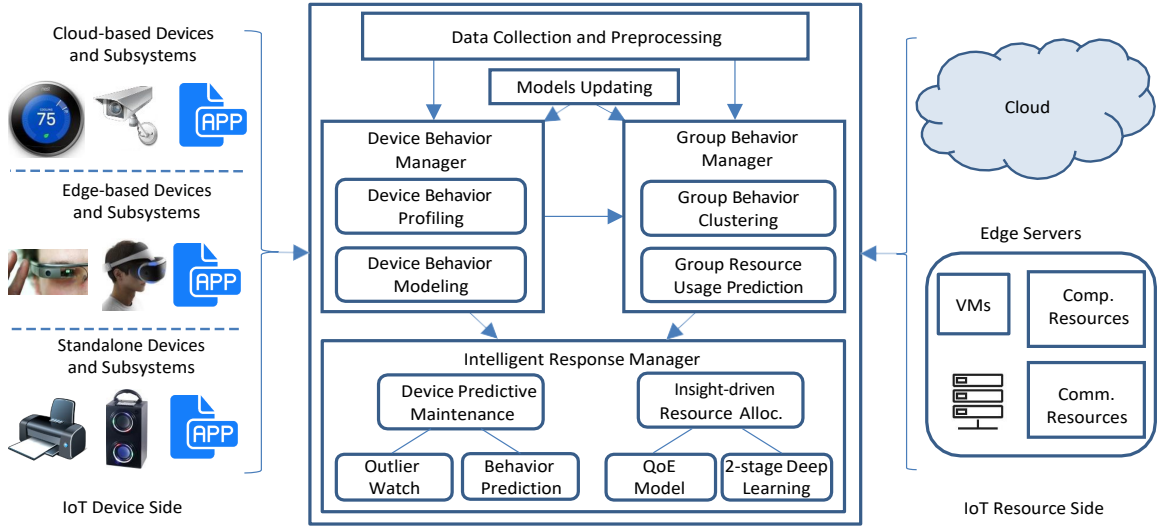

Fig. 3: ORCA architecture.

*2) "Synthesizing": Owner-centric Group and Subsystem Behavior Modeling:* This step focuses on the IoT owners' application and resource manager roles, and aims to synthesize group and subsystem level behavior and resource usage insights for the following management decision-making.

i) Group/Subsystem Behavior Synthesizing. This stage aggregates individual devices' behavior scores, synthesize group behavior, and identify outliers. We allow the IoT owners to flexibly define "groups" as IoT subsystems (e.g., all the IP cameras for a video surveillance subsystem), locations (e.g., all devices on a specific floor of a building), or device batches (e.g., all devices procured in the same batch with similar software and hardware configurations). With such grouping, we can run clustering to synthesize group insights and identify potential "outliers", and help owners find out the device numbers with lowest or highest scores, and the current and history score distributions for IoT subsystems. These insights can help understand system dynamics and identify problematic devices. It can help owners find out whether the devices are with lower scores than other areas for a location/network. These insights can help identify large-scale botnet activities or malfunction caused by faulty IoT gateways. It can also help decide whether they are showing similar low scores for a device batch. These insights can help identify batches that need updates or attention.

ii) Group/Subsystem Resource Usage Prediction. This stage synthesizes how the shared edge resources are currently being or will be used by groups or subsystems. For example, if a video cognitive subsystem uses 90% of the total edge resource while contributing only 10% of the financial revenue, IoT owners may be motivated to rebalance the resource budgets among subsystems. Similarly for abnormal cases when the group resource usage with certain locations/networks or device batches reaches alarming ranges, it may justify certain management actions. In addition, if the group resource usage insights are combined with accurate and prompt prediction, they can help IoT owners to efficiently allocate the shared edge resources to the different subsystems, better prepare for large-scale abnormal incidents, and protect the overall welfare of these systems. A typical candidate tool is LSTM.

*3) "Responding": Owner-centric and Data-driven Management Responses:* This step takes IoT owners' all roles as device, application, and resource manager, and aims to enable the IoT owners to make well-informed and intelligent management decisions at edge to manage individual devices and the shared edge resources among all subsystems.

i) Device-level Predictive Maintenance. The IoT devices' behavior can be very dynamic. To identify devices that need future maintenance with confidence, we conduct behavior prediction over the group outliers identified in the above second step. Behavior prediction in heterogeneous Edge-IoT needs to handle multiple behavior models, large amounts of historical records, random behavior pattern changes with stationary and non-stationary distributions, varied prediction window length for different reaction delays, and fast and cost-efficient prediction.

ii) Intelligent Edge Resource Allocation. With the obtained insights, we then aim at intelligently allocating edge resource to jointly optimize user experience and resource utilization, while containing bad behavior. We design a novel resource allocation scheme that does two things. First, we build a new QoE model to quantify the devices' satisfaction, which comprises heterogeneous subsystems' QoS requirements and priorities. The model accounts for device current behavior, predicted behavior, group behavior, and predicted edge resource usage. Second, we build a novel two-stage deep online learning [15] scheme to jointly optimize user experience and edge resource usage across subsystems.

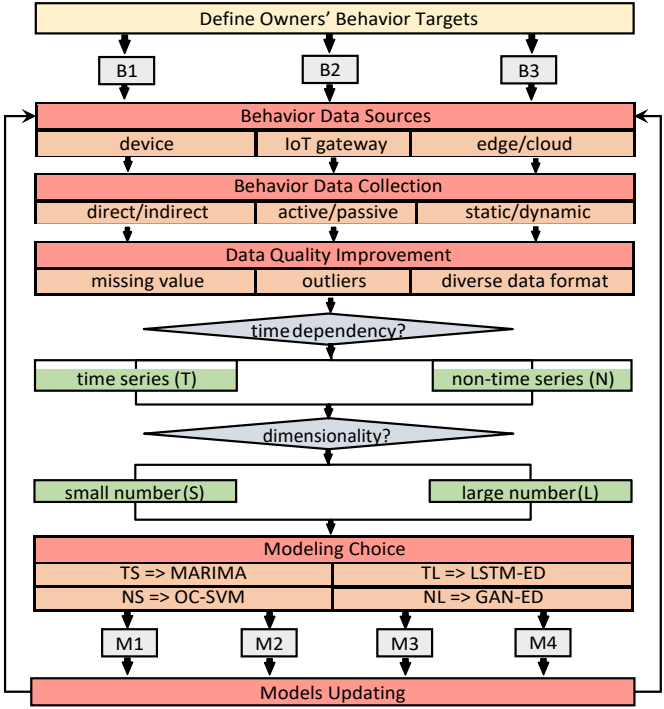

Fig. 4: Device behavior profiling and modeling workflow.

## V. Evaluation and Discussions

In this section, we conduct preliminary evaluations on scalability both qualitatively and quantitatively.

### A. Qualitative Evaluation and Discussions

First, for the vendor/SP-centric paradigm, the vendor like Google or Amazon may have to manage in central locations (clouds) for large numbers of devices resided in many owners' domains. In comparison for ORCA, in each owner's domain, the number of devices they own and manage is much less. Second, ORCA incorporates scalability and extensibility supports. As illustrated in Fig. 2 in Section III.B, in ORCA, the required number of trained and deployed behavior models is approximately $O(M \cdot N)$. It is decided by the number of device types $N$ (such as cameras and drones) and the number of behavior models $M$ (B1 to B4) for each device type. With such design, the overall model cost of the ORCA will increase polynomially, regardless the possible exponential increase of the number of devices. Thus, qualitative speaking, when a new device type comes to the market, ORCA will be able to scale up efficiently with the IoT device variety $N$. When the owner wants to extend his/her management interests, ORCA will be able to scale up by increasing new management interests $M$. In addition, to balance the cost (training/operation) and performance (speed/accuracy) for the IoT behavior models, ORCA provides four candidate modeling approaches for the owners to choose based on the feature dimensionality and data time dependency.

### B. Quantitative Evaluation and Discussions

We also present a quantitative evaluation on scalability. We build a testbed with several typical IoT devices such as IP cameras and temperature sensors, and an ORCA manager using a Raspberry Pi 3 with 1.4GHz CPU, 1GB RAM, and 16GB storage. The behavior dataset consists of time-series (TS) and non-time-series (NTS) samples. TS samples are collected every 30 minutes where each sample is a single-variate sequence with 90 data points. NTS are sampled once per minute where each sample has 80 features from B1 to B4. We evaluate behavior modeling cost by implementing the four models proposed in Section IV.B, where GAN-ED and LSTM-ED have two neural layers with 64 and 32 neurons in both encoder and decoder networks. For NTS, behavior measuring tasks are uniformly distributed in one minute. With these configurations, we observe the following results. First, the size of one TS sample and one NTS sample are about 1KB and 0.5KB respectively. Suppose that the IoT devices number is 120 in the service coverage of one manager, the total behavior data size for all devices are 60KB per minute for TS and 120KB every 30 minutes for NTS. Second, we observe the key scalability parameters including models' sizes, running time, and running memory in Table II. For the model size, LSTM-ED is the biggest among others and each one takes only 630KB. For the running time, the one-time behavior evaluation takes at most 233 ms, which meets the responding latency requirement of ORCA. For the running memory, we observe the run-time memory consumption of any model does not exceed 184 MB. The evaluation results demonstrate that all the models can run on the ORCA server with very limited resource cost. In our experiment settings, a resource-constrained Raspberry Pi-based ORCA manager can manage a fair number of devices while causing limited network traffic and storage overhead.

TABLE II: Costs of the four models.

| Model | Model Size (KB) | Model Running Time (ms) | Model Running Memory (MB) |
|---|---|---|---|
| OC-SVM | 103 | 6 | 97 |
| MARIMA | 36 | 12 | 84 |
| GAN-ED | 345 | 62 | 163 |
| LSTM-ED | 630 | 233 | 184 |

## VI. Conclusions

It is a significant challenge to manage massive numbers of diverse devices and heterogeneous Edge-IoT applications. The current methods mostly manage these systems as separate vertical "silos", or in a vendor/SP-centric way, which suffers from a series of limitations. To address the challenges, in this paper, we proposed a new owner-centric paradigm named ORCA empowered by data-driven approaches and machine learning techniques. ORCA aims to fill the gap and complement the missing pieces of the existing management approaches. It provides a scalable and extensible framework for IoT asset owners to perform data-driven 3-step management to complete the the "observing, synthesizing, and responding" management cycle. The preliminary evaluation has demonstrated the effectiveness of the proposed ideas. Our future work include further validation and integration of the building pieces.

## Acknowledgment

The work is supported by NSF research grant No. 1909520.

**Jianli Pan** is currently an Associate Professor in the Department of Computer Science at the University of Missouri, St. Louis, MO USA. He obtained his Ph.D. degrees from the Department of Computer Science and Engineering of Washington University in St. Louis, USA. He is an associate editor for both IEEE Communication Magazine and IEEE Access. His current research interests include Internet of Things (IoT), edge computing, machine learning, and cybersecurity.

**Jianyu Wang** is currently a Ph.D. student with the Department of Computer Science at the University of Missouri, St. Louis. He received an M.S. in Electrical and Computer Engineering from the Rutgers University, New Brunswick. His current research interests include edge cloud and mobile cloud computing.

**Ismail AlQerm** is a postdoctoral research associate in the Department of Computer Science at University of Missouri, St. Louis. He received his Ph.D. in computer science from King Abdullah University of Science and Technology (KAUST) in 2017 and was among the recipients of KAUST Provost Award. His research interests include edge computing, IoT resource allocation, machine learning in wireless networks, and software defined radio.

**Yuanni Liu** is an associate professor at the Institute of Future Network Technologies, ChongQing University of Posts and Telecommunications, China. She received her Ph.D. from the Department of network technology Institute, Beijing University of Posts and Telecommunications, China, in 2011. Her research interests include mobile crowd sensing, IoT security, and data virtualization.

**Zhicheng Yang** received his Ph.D. degree in Computer Science at University of California, Davis, CA, USA, in 2019. He is currently a senior research scientist at PAII Inc, USA. His current research interests include millimeter wave sensing, 60 GHz communications and networking, and mobile computing.